# Local photo-mechanical stiffness revealed in gold nanoparticles supracrystals by ultrafast small-angle electron diffraction

Giulia Fulvia Mancini[1,2,*], Francesco Pennacchio[3], Tatiana Latychevskaia[4], Javier Reguera[5], Francesco Stellacci[6] and Fabrizio Carbone[3,*]

[1] Laboratory for Ultrafast Spectroscopy, Lausanne Center for Ultrafast Science (LACUS), École Polytechnique Fédérale de Lausanne, CH-1015 Lausanne, Switzerland.

[2] Paul Scherrer Institut, WSLA/210, 5232 PSI Villigen, Switzerland.

[3] Laboratory for Ultrafast Microscopy and Electron Scattering, Lausanne Center for Ultrafast Science (LACUS), École Polytechnique Fédérale de Lausanne, CH-1015 Lausanne, Switzerland.

[4] Physics Institute, University of Zurich, Winterthurerstrasse 190, 8057 Zurich, Switzerland.

[5] CIC biomaGUNE, Paseo de Miramón 182C, 20009 Donostia-San Sebastian, Spain. Ikerbasque, Basque Foundation for Science, 48011 Bilbao, Spain.

[6] Supramolecular Nanomaterials and Interfaces Laboratory, Institute of Materials, École Polytechnique Fédérale de Lausanne, CH-1015 Lausanne, Switzerland.

[*] Email: giulia.mancini@psi.ch; fabrizio.carbone@epfl.ch

**Abstract**

We demonstrate that highly-ordered two-dimensional crystals of ligand-capped gold nanoparticles display a local photo-mechanical stiffness as high as that of solids such as graphite. In out-of equilibrium electron diffraction experiments, a strong temperature jump is induced in a thin film with a femtosecond laser pulse. The initial electronic excitation transfers energy to the underlying structural degrees of freedom, with a rate generally proportional to the stiffness of the material. With femtosecond small-angle electron diffraction, we observe the temporal evolution of the diffraction feature associated to the nearest-neighbor nanoparticle distance. The Debye-Waller decay for the octanethiol-capped nanoparticles supracrystal, in particular, is found to be unexpectedly fast, almost as fast as the stiffest solid known and observed by the same technique, *i.e.* graphite. Our observations unravel that local stiffness in a dense supramolecular assembly can be created by Van der Waals interactions up to a level comparable to crystalline systems characterized by covalent bonding.

**Main text**

Two-dimensional supracrystals created by self-assembly of nanoparticles (NPs) offer a route toward engineering materials with specific functionalities [1,2]. The building blocks are metallic core NPs functionalized with a ligand shell of organic molecules, determining the macroscopic structural, electronic, optical and magnetic properties of these systems [3–6]. In alkanethiol-protected gold NPs supracrystals, the competition between thermodynamic driving forces can lead to structural phases ranging from crystalline to glassy by simply changing the length of the alkanethiol molecules [7]. The structural diversity and the stabilization of these supracrystals are primarily affected by the level of ligand interdigitation [8,9].

The simultaneous characterization of the NP cores and their ligands shell, as well as of the short-range properties of the NPs assembly, has proven challenging so far [1]. Mueggenburg *et al.* showed that monolayers of dodecanethiol-capped gold nanoparticles display remarkable properties of mechanical strength, comparable to that of glassy polymers, accompanied by robustness and resilience at higher temperatures [10]. In this work, however, macroscopic (*i.e.*, long range) properties of the overall colloidal crystal were analyzed.

In alkanethiol-capped NPs colloidal crystals, the local (*i.e.*, short-range) mechanical properties are primarily affected by the local arrangement of NPs cores and organic ligands within single grains. For this, a scattering technique is required, which can combine Ångstroms (Å) spatial resolution with sensitivity to light elements. Effects of ligand chain length on NPs packing density or disorder have recently been explored in two-dimensional ligand-stabilized NPs supracrystals by Kim *et al*. [7], where equilibrated monolayers were produced by cyclic compression and relaxation in Langmuir trough, and their phase transition from crystalline to liquid through a hexatic phase was interpreted as entropy-driven phenomenon associated with steric constraints between ligand shells.

Only recently, ultrafast small-angle electron diffraction has enabled to resolve both their static ordering properties and their photo-induced motions [11] with combined femtosecond (fs) temporal and Ångstroms (Å) spatial resolution. In this work, we observe that the light-induced decay of the intensity of the diffraction feature associated to the local (nearest to next-nearest neighbor) hexagonal arrangement of the gold NPs in each supracrystal depends on the functionalizing ligands length. For the shortest ligands, we find that the rate of this decay is as fast as what is observed in a very stiff solid such as graphite, characterized by strong homonuclear covalent bonding. The transient response from supracrystals of dodecanethiol-coated gold nanoparticles, instead, is found to be significantly slower, comparable to softer systems. Our experimental results are supported by simulations which demonstrate that the local symmetry of the NPs within the supracystal grains affects the short-range degree of coupling between the electronic and lattice degrees of freedom.

We conducted experiments on three different 2D nanoparticles supracrystals. Each sample is consisting of monodisperse gold NPs of the same size (~5 nm) coated with different alkanethiols (R–SH, R = $C_nH_{2n+1}$): 1-octanethiol (n=8), 1-dodecanethiol (n=12) and 1-octadecanethiol (n=18). Within the text, we will refer to these samples as C8, C12 and C18, respectively. Each NPs supracrystal monolayer was obtained by Langmuir−Schaefer deposition [11–13] (details provided in the Supplemental Material). We generated electrons at 30 keV with a frequency tripled ultrafast 800 nm laser, 50 fs pulse duration, 20 kHz repetition rate. High-brightness, ultrashort (<300 fs) bunches were obtained by radiofrequency compression and were focused with a set of magnetic lenses to a ≈160 μm spotsize onto each supracrystal [14]. Impulsive photoexcitation of each sample is obtained with 1.5 eV light pulses with ≈220 μm spotsize at the sample plane. The supracrystals time-dependent response is probed with small-angle ultrafast electron diffraction in transmission geometry [11], with the diffraction patterns forming onto a phosphor screen and detected by a single-electron counting charge-coupled device (Fig. 1).

Structure retrieval methods based on the calculation of the angular Cross-Correlation Function (CCF) have proved effective in retrieving the local arrangement and the symmetry of the NPs in the supracrystal [11]. In our experiment, this information is contained in the small-angle region of the diffraction patterns from each supracrystal, at the scattering vector $s_1$ marked in the inset of Fig. 1. For each sample, the diffraction feature at the scattering vector $s_1$ is related to the real space distance $d_1$ of crystallographic planes created by nanoparticles in the supracrystal (Fig. 2b). The values of $s_1$ for each sample are reported in Table 1. As mathematically derived in Ref. [15], characteristic symmetries in such dense systems of identical particles can be detected in the CCF when the coherence length of the probing wave is at least comparable to the size of the single particle. The normalized CCF is defined as [16–21]:

$$C_{norm}(\Delta) = \frac{\langle I(s_1,\varphi)I(s_1,\varphi+\Delta)\rangle_\varphi - \langle I(s_1,\varphi)\rangle^2_\varphi}{\langle I(s_1,\varphi)\rangle^2_\varphi}, \qquad (1)$$

where $I(s_1,\varphi)$ represents the scattered intensity at the scattering vector $s_1$ and the angle $\varphi$; $\Delta$ is the shift between the two angles, (inset of Fig. 1) and $\langle\ \rangle_\varphi$ denotes an averaging over $\varphi$. Following the approach described in Ref. [11], the normalized CCF at the scattering vector $s_1$ for each sample was obtained from the one-dimensional Fourier spectrum of the scattered intensity $I(s_1,\varphi)$. Details on this analysis are available in the Supplemental Material.

Fig. 2a displays the CCFs retrieved at equilibrium (i.e. before photoexcitation) for the C8 (red), C12 (orange) and C18 (pink) supracrystals. The 6-fold modulation of the CCF from the C8 and C12 samples reflects the hexagonal close-packed arrangement of the NPs in the supracrystals (Fig. 2b). This suggests the presence of a crystalline structural phase in which neighboring NPs are held together within single grains by attractive Van der Waals forces that lead to the favorable interdigitation of the ligands [7–9]. Thus, NPs in each grain arrange in crystallographic-like planes with distance $d_1 = \frac{2\pi}{s_1}$ (Fig. 2b). The absence of recognizable symmetries for the C18 sample indicates a lack of short-range order in the NPs self-assembly, likely due to repulsive forces winning over Van der Waals attractive interactions [7], as well as to lower NPs solubility and mobility at room temperature. A simulation of a single alkanethiol-coated NP is reported in Fig. 2d. Gold NPs cores of ~5 nm are simulated as having a polyhedral morphology, with the gold atoms arranged in a face-centered cubic (*fcc*) lattice [11,22]. The chemical structure of each ligand (n = 8, 12, 18) is displayed for clarity. The ligand lengths and the average core-core NPs distances for each supracrystal, retrieved experimentally with electron diffraction, are summarized in Table 1.

We analyzed the transient changes in the NPs hexagonal arrangement at $s_1$ for the C8 and the C12 supracrystals and compared their time responses (Fig. 2c). The radial average intensity at $s_1$ is calculated as:

$$I(s_1) = \frac{1}{2\pi} \int I(s_1, \varphi) \mathrm{d}\varphi. \tag{2}$$

The transient change of $I(s_1)$ for the C8 (red circles) and the C12 (orange squares) supracrystals is reported in Fig. 2c. Each intensity trace was normalized to the average value at negative times ($t<t_0$) and fitted to a mono-exponential curve (solid lines). Photoinduced thermal disorder in the NPs hexagonal arrangement is evidenced in both samples by the transient decrease of $I(s_1)$. Remarkably, the $I(s_1)$ decay time-scale for C8, τ = 2.6 ± 0.3 ps, is significantly shorter than the one for C12, τ = 12.1 ± 0.9 ps.

The $I(s_1)$ suppression for C8 and C12 is due to the energy transfer between the electronic excitation made by light and the underlying structural degrees of freedom of the supracrystal. In Fig. 2c, both decay traces are compared to the ones detected in transmission ultrafast electron diffraction in two solid state systems characterized by a vastly different electron-phonon coupling strength, namely graphite [23] (purple asterisks) and Bismuth [24] (grey crosses). Intuitively, in nanostructured systems such as the ones investigated in this work, the presence of attractive Van der Waals interactions among ligands should act as a glue to hold the NPs in the supracrystal together *via* interdigitation. Such non-covalent "bonding", and its dynamical response to energy transfers, should lead to a disordering of the NPs local arrangement on time-scales comparable if not slower to those of a soft solid, such as Bismuth. While our observations suggest that this scenario applies for the C12 supracrystal, where the relaxation follows a timescale τ = 12.1 ps, the dynamics observed in the C8 supracrystal is dramatically different and unexpected.

In the C8 supracrystal, the intensity drop of the diffracted beam follows a Debye-Waller factor comparable to the one of graphite, which is to date the fastest ever observed in ultrafast electron diffraction [23,25,26]. In graphite, which microscopic structure is characterized by the presence of strong homonuclear covalent bonds, a bi-exponential decay of $I(s_{110})$ revealed the presence of strong coupling between the electronic subset with a small subset of lattice degrees of freedom (τ = 250 fs), followed by carriers cooling through electron-phonon and phonon-phonon scattering with a timescale τ = 6.5 ps [23]. The comparable rapidity in the suppression of the diffracted intensity for C8 suggests that in this supracrystal the interdigitation of the shorter ligands provides a very efficient channel for transferring energy between the initial electronic excitation to structural motions of the NPs.

To rationalize this result, one may consider that upon photoexcitation of a metallic system, the evolution of the electronic temperature and lattice temperature can be described via the popular two temperature model [27], in which hot electrons exchange energy with the phonons sub-system. Consequently, such a sub-system increases its temperature. The speed of this energy transfer is directly reflected by the evolution of the Debye-Waller effect in a diffraction experiment, as phonon excitation disorders atomic position, thus affecting the diffraction intensity [25,28,29]. In more complex solids, it is possible that only a sub-set of phonons is excited first [25,29–31], or that other bosonic subsystems can drain the excess energy from the hot electrons. In cuprates superconductors, for example, a four-temperature model was proposed to account for the energy transfer between hot electrons, phonons, and spin fluctuations [32,33]. Despite the rich variety of circumstances that one can encounter, it is always true that the speed of the electronic or lattice temperature decays is directly related to both the coupling between the initial excitation and the modes transferring the excess energy and their own energy.

Contrarily to all systems described above, the structural modes of a self-assembled supracrystal are not an established concept. However, in the case of the gold nanoparticles assemblies investigated in this work, it is reasonable to assume that the only subsystem to which the initial electronic excitation can transfer energy to is the supracrystal structure itself. In fact, as demonstrated in [7], an increase in ligand length limits the possible configurations for the ligand chains interdigitation, yielding and increase in conformational disorder. As such, these systems qualify as disordered elastic media in which the ligands are the non-covalent «springs» which couple the gold cores (*i.e.*, the «masses»), while their length affects their interpenetration, *i.e.* «the spring constant». Hence, the transient decay of the Debye-Waller factor we observe is a signature of a very efficient way of transferring energy from the light-induced excitation and the nanoparticle supracrystal. Since our probe is the local order of such a supracrystal, *i.e.* the nearest-neighbor environment of each nanoparticle, it is reasonable to assume that at that level one can expect to have similar characteristic vibrations as those of a mass coupled to another one via a spring. In this scenario, the only plausible pathway to efficiently transfer energy from the electronic excitation of the nanoparticles to the supracrystal is if the spring constant is very high.

We point out that the microscopic description of the collective modes responsible for such an energy transfer between the electronic excitation and the supracrystal, supporting our assumptions, is subject of current investigation. In fact, a complete theoretical framework is still lacking to date, despite increasing interest demonstrated –for example- by the postulated existence of short-range phonons in soft matter and liquids, also termed anakeons [34].

The photoinduced disorder observed in both C8 and C12 supracrystals is accompanied by annealing of NPs grains [11], evidenced by a transient increase in the signal-to-noise ratio of the CCFs at different time delays (see Supplemental Material). This behaviour indicates that local order is impulsively triggered, within the relevant time-scales, in supracrystals where the NPs distribution is stabilized by ligand interdigitation.

Next, we demonstrate for each sample the presence of a direct correlation between the observed Debye-Waller decay, and the symmetry of the NPs within the supracrystal grains. Ligand length-dependent order-disorder correlations in the three supracrystals were explored in a series of simulations. The Transmission Electron Microscopy (TEM) images of the C8, C12 and C18 samples measured with UED are displayed in Fig. 3. The corresponding diffraction patterns are simulated as the squared amplitude of the Fourier Transform (FT) of the TEM images (see Supplemental Material). The patterns in Fig. 3 directly visualize a hexagonal order at $s_1$ in the C8 and C12 samples. The core-core distance values extracted from the FTs of the TEM images are in agreement with the values retrieved experimentally with electron diffraction (Table 1). For the C18 supracrystal, the observed Airy pattern is created by the diffraction on randomly distributed objects, i.e. the NPs, with an aperture diameter equal to the diameter of the NP. The Airy pattern indicates a glassy structural phase of the C18 supracrystal which was experimentally demonstrated with CCF analysis (Fig. 2a). The average of the NPs in C18 was estimated to 5.4nm by fitting the radial averaged intensity $I(s)$ with the following function:

$$I(\rho) = \left(\frac{J_1\left(\frac{\pi\rho d}{\lambda z}\right)}{\left(\frac{\pi\rho d}{\lambda z}\right)}\right)^2 \qquad (3)$$

With $J_1$ Bessel function of the first kind, $d$ the NPs diameter, $\rho$ the abscissa coordinate, $\lambda$ the wavelength and $z$ the sample-detector distance.

Sphere lattice model (SLM) simulations are used to assign the different local distribution of the NPs within supracrystal grains (Fig. 4). The models are created from a two-dimensional lattice of opaque spheres (~5 nm diameter) arranged in a hexagonal lattice. The program selects round domains and rotates them by the angle η, where η is Gaussian distributed with the mean = 0 and standard deviation σ. The domains are selected to have a size of 60nm and a centre-to-centre distance between domains of 80nm. Two distributions are shown in Fig. 4: (i) σ = 10° with a perfectly ordered spheres arrangement preserved within each domain (left), and (ii) σ = 10°, with disorder of individual NPs positions introduced by adding a random shift up to Δr = ±1 nm (right). The location of $s_1$ is indicated in Fig. 3 for the experimental and Fig. 4 for the simulated Fourier transforms. In each, $s_1$ marks the first order of diffraction from the crystallographic planes with distance $d_1$ displayed in Fig. 2b.

The similarity between (i) the [10°, 0] model with the C8, C12 samples, and (ii) the [10°, ±1 nm] model with the C18, is further confirmed by the radial averaged intensity plots in Fig. 5(a, b). Multiple orders of diffraction from the $s_1$ crystallographic planes are observed in systems characterized by the presence of local symmetry (C8, C12). These diffraction orders progressively smear into a pattern from cahotically distributed objects by increasing the randomization in the NPs positions (C18). In the C8, C12 supracrystals where a transient suppression of $I(s_1)$ is observed, the simulations confirm a direct connection between the observed photo-mechanical local stiffness with the presence of a recognizable translational symmetry of the NPs distribution within single domains. In C18, where no local symmetry is found, the randomized distribution of the scattering objects within each domain leads to the presence of a liquid-like phase, where no correlation between transient changes and NPs arrangement can be unraveled. These observations are supported by the analysis of the CCFs retrieved from the Fourier transforms of the TEM images in Fig. 5c. We point out that minor differences (for example, in contrast at $s_1$) between the FTs from TEM experimental images (Fig. 3) and the FTs from the corresponding SLM simulations (Fig. 4), are solely due to a different degree of disorder of spheres, and to the lack of random additional noise in the SLM simulated data, as confirmed by the full agreement between the CCFs at $s_1$ from experimental data (Fig. 2a) and FTs of TEM experimental images (Fig. 5c).

In this work, we reported that local mechanical stiffness can be created in supramolecular assemblies by Van der Waals forces to an extent comparable to systems characterized by strong covalent bonding. The key to our observation is a powerful technique which combines small-angle electron diffraction (Å spatial resolution) with ultrafast (fs) temporal resolution. Moreover, we demonstrate that the local symmetry of the NPs within the supracystal grains affects the short-range degree of coupling between the electronic and lattice degrees of freedom. We point out that our small-angle scattering technique is sensitive to the short-range arrangement of a few nearest-neighbors (alkanethiol-capped gold NPs) within crystal grains, in samples that are overall characterized by the presence of discontinuities, as dislocations and grain boundaries. Thus, the transient response observed in each sample unveils information about photo-mechanical local stiffness, *i.e.* at the nearest/next nearest neighbor level. Because of the presence of defects and dislocations in the overall colloidal crystal, we remark that the observed stiffness is not expected to directly translate into a macroscopic mechanical property. Our results provide a seed for new theoretical models of structural collective modes in soft-matter systems.

**Acknowledgements**

This work was supported by the Swiss National Science Foundation (SNSF) through the grant no.PP00P2-128269/1. G.F.M., F.C. and F.S. designed research; G.F.M. and F.P. conducted the experiment; G.F.M. and T.L. analyzed the data and performed simulations; J.R. prepared the samples. All authors commented on the results and on the manuscript. The authors declare no competing financial interest.

**Figures**

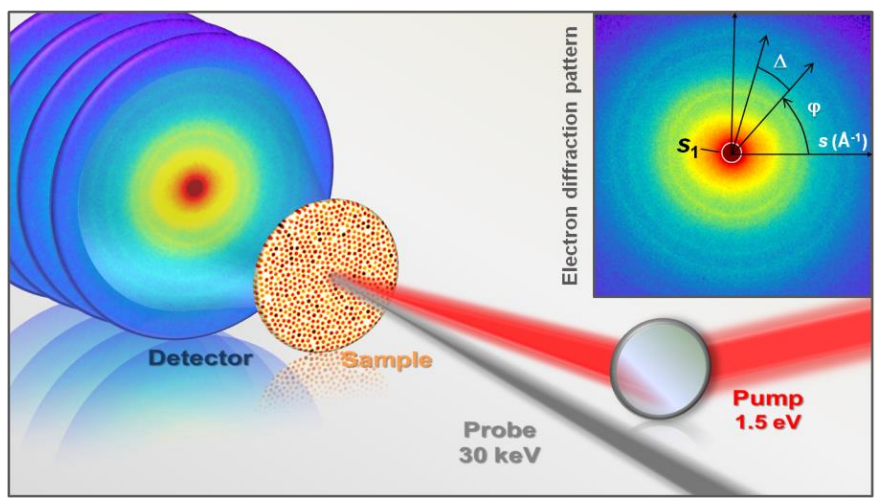

FIG. 1. Ultrafast small-angle electron scattering experimental layout. Radiofrequency-compressed ultrashort and ultrabright electron bunches are focused onto each supracrystal. Transient changes in the sample are initiated by 1.5 eV light pulses. At each time delay, the electrons scattered from the sample are collected on an intensified electronic imaging charged-coupled device (CCD) capable of single electron detection. The inset shows a typical diffraction pattern recorded in the experiment and the concept of angular cross-correlation analysis. The Debye-Scherrer rings from polycrystalline gold are detected at large angle. The information related to the NPs symmetry in each supracrystal is included at small angle. The NPs static arrangement and their ultrafast dynamics are unraveled in our experiment by angular cross-correlation analysis.

| Sample | $s_1$ [Å$^{-1}$] | Ligand length [nm] | Core-core distance [nm] | Time scale $T$ [ps] |
|---|---|---|---|---|
| C$_8$ | 0.118 | 1.285 | 6.1 | 2.6 ± 0.3 |
| C$_{12}$ | 0.095 | 1.789 | 7.2 | 12.1 ± 0.9 |
| C$_{18}$ | 0.098 | 2.542 | 7.4 | – |

TABLE 1. Relevant distances and time-scales for the C8, C12 and C18 supracrystals.

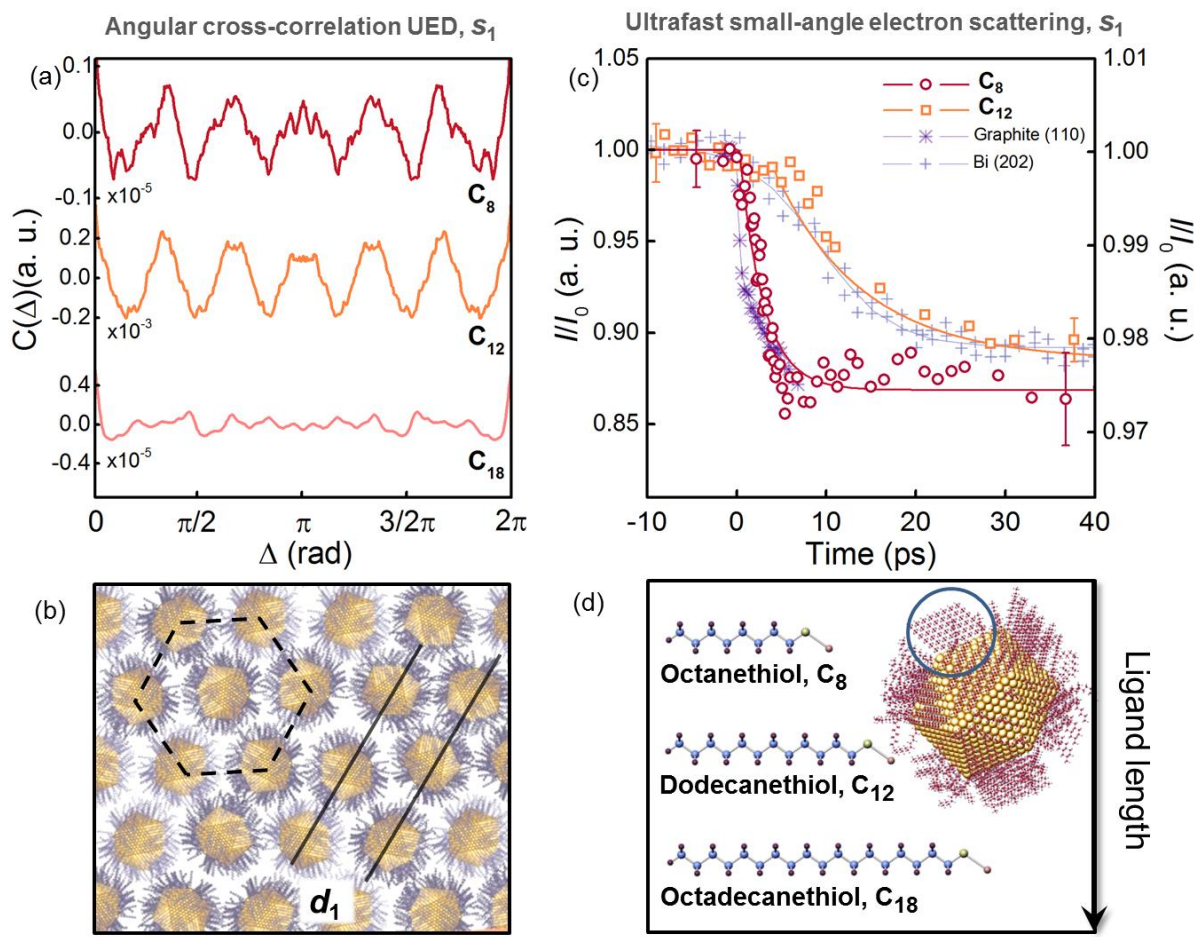

FIG. 2. Ligand length-dependent dynamic response of C8 and C12 supracrystals upon photoexcitation. (a) The degree of hexagonal symmetry of the NPs within the supracrystal is unraveled by angular cross-correlation analysis. The electron diffraction data show a clear hexagonal arrangement of the NPs in the C8 (red) and C12 (orange) supracrystals, whereas a disordered, liquid phase is found for the C18 (pink) sample. (b) Schematic illustration of a two-dimensional monolayer of alkanethiol-functionalized NPs. The NPs are arranged in a hexagonal lattice and create crystallographic planes which distance $d_1$ is primarily a function of the ligand length. (c) NPs dynamics for the C8 and C12 supracrystals are compared to the time-response of graphite and Bismuth in UED transmission experiments. The C8 supracrystal (red circles) shows a decay in diffracted intensity comparable to graphite (purple stars), suggesting a strong coupling of the electronic and lattice degrees of freedom. When the ligand length increases (C12, orange squares) the system is characterized by a softer electron-phonon coupling, similar to soft metals such as Bismuth (grey crosses). Graphite data were digitized from [23]. Bismuth data digitized from [24]. (d) Simulation of a single thiolated NP and comparative display of ligand structures and length (reported in Table 1).

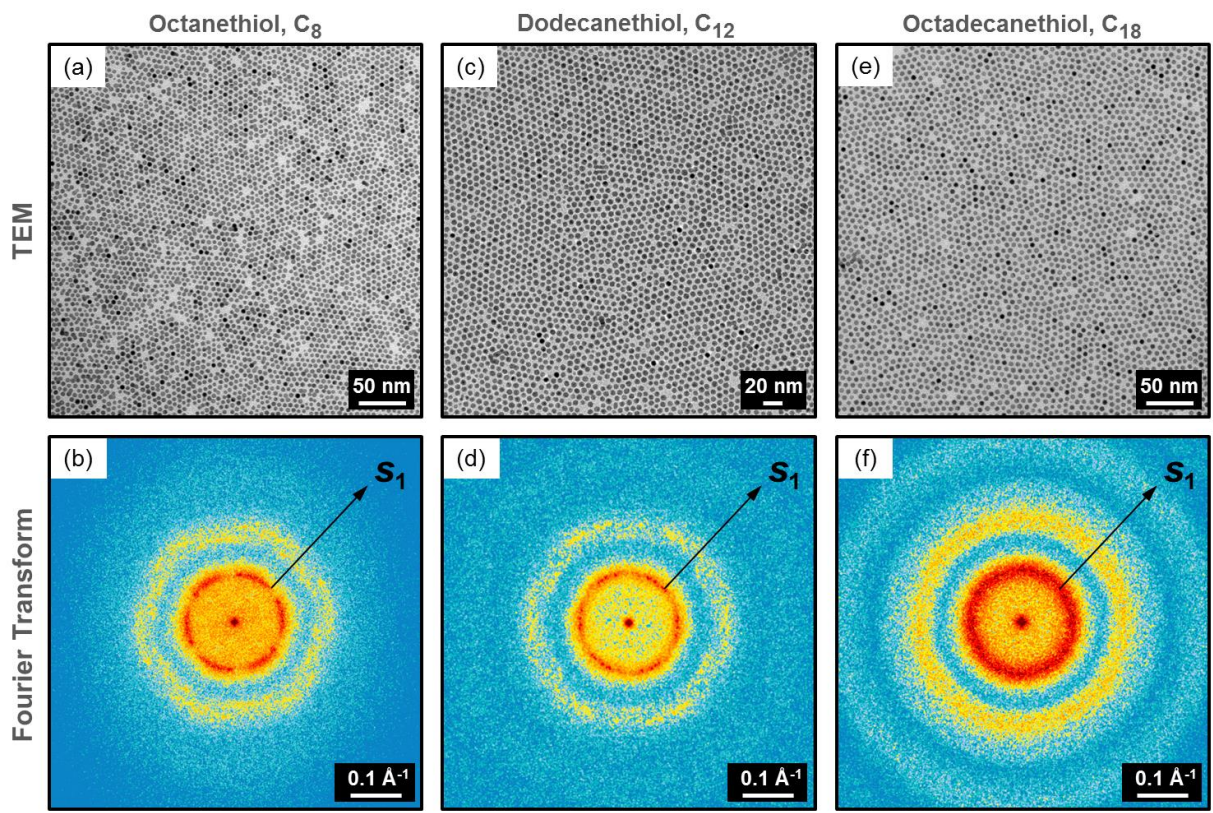

FIG. 3. TEM images and Fourier transforms. (a, c, e) TEM images of the C8, C12 and C18 supracrystals (left to right) measured with UED. (b, d, f) Diffraction patterns simulated as the squared amplitude of the Fourier Transform of the TEM images in (a, c, e) respectively, confirming the increasing disorder as a function of the ligand length (left to right). In each simulation, $s_1$ marks the first order of diffraction from the crystallographic planes with distance $d_1$ displayed in Fig. 2b. Each two-dimensional Fourier transform is normalized to the number of nanoparticles detected in the corresponding TEM image.

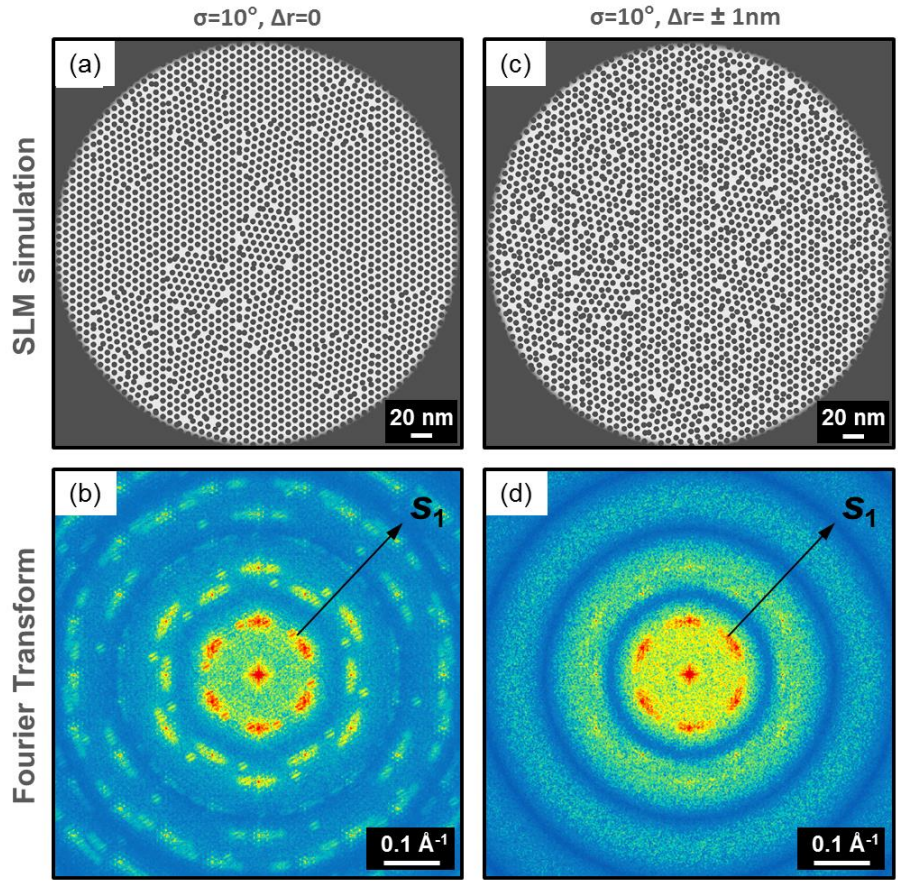

FIG. 4. Supracrystal simulations and Fourier transforms. Simulations of the TEM images considering each NPs as a spherical opaque object. (a) σ = 10° with a perfectly ordered spheres arrangement preserved within each domain. (b) Diffraction patterns simulated as the squared amplitude of the Fourier Transform of (a). (c) σ = 10°, with disorder of individual NPs positions introduced by adding a random shift up to Δ$r$ = ±1 nm. (d) Diffraction patterns simulated as the squared amplitude of the Fourier Transform of (c). Each two-dimensional Fourier transform is normalized to the number of nanoparticles in the corresponding SLM simulation.

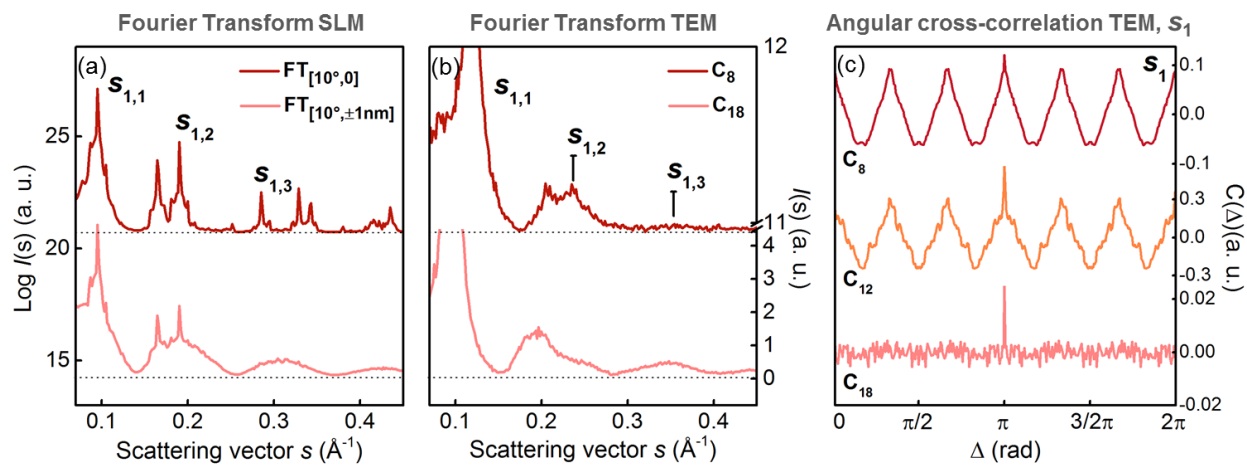

FIG. 5. Compared intensity profiles from experimental and simulated Fourier Transforms and angular cross-correlation analysis of the C8, C12, C18 TEM images. (a) Radial averaged intensity of the Fourier Transform of the [10°, 0] simulation (red) and of the [10°, ± 1 nm] simulation (pink). (b) Radial averaged intensity of the Fourier transform of the TEM images for C8 (red) and C18 (pink). (c) Angular cross-correlation functions at $s_1$ from the Fourier Transform of the TEM images of the three samples. The location of $s_1$ in the two-dimensional simulation is reported in Fig. 3 (b, d, f). For each curve, $s_1$ refers to the first order of diffraction from the crystallographic planes with distance $d_1$ displayed in Fig. 2b. The angular cross-correlation reflects the degree of disorder in the hexagonal distribution of the NPs in the sample, and it is in agreement with the CCFs extracted with small-angle electron diffraction.

# Supplemental Material

## EXPERIMENTAL DESIGN

The driving laser is a KMLabs Wyvern centered at a wavelength of 800 nm, with 0.65 mJ pulse energy, 50 fs pulse duration, at a repetition rate of 20 kHz. The shot-to-shot noise on the energy per pulse in the experiment is below 0.2%. Electrons are generated in a DC gun by back illumination of a silver-coated sapphire window on a high voltage (30 kV) photocathode with 266 nm light pulses. UV light is obtained from the fundamental driving laser by second and third harmonic generation. The photoemitted electron bunches are directed to the sample with a collimating solenoid lens ($I$ = 25 A) and a focusing solenoid ($I$ = 1.72 A). Their temporal chirp is compensated with a radiofrequency cavity operating in the $TM_{010}$ mode [14, 35]. This results in ~300 fs/160 µm bunches at the sample, each containing up to 6 × $10^5$ electrons. The diffraction pattern was formed on a phosphor screen and detected on a PI-MAX:1300 HB intensified electronic imaging camera capable of single electron detection with an array of 1300 x 1340 pixels (side length of 20 µm). The collected diffraction patterns were single binned on-chip at 1 MHz pixel readout rate. The experiments were conducted with consistent parameters for all three samples. At every time delay, a single image is the result of 500 accumulations with 300 gates per exposure. The diffraction patterns were acquired an overall current of 320.4 pA for every time delay on each sample. The total charge amount, distributed over several electron pulses, was calculated taking into account for the number of accumulations/time-delay and the total number of time delays, and was estimated to: 0.25 µC, 0.411 µC and 0.23 µC for the $C_8$, $C_{12}$, $C_{18}$ samples respectively. Photoinduced changes in the samples were initiated by 1.5 eV pump-pulses focused to a spot of 220 µm. The incident fluence on the sample was 13 mJ/cm$^2$ (100 mW at 20 kHz with 220 µm spotsize). The effectively absorbed fluence was estimated around 100 µJ/cm$^2$, based on the optical reflectivity of gold in a layer of 7 nm thickness [36], the sample density, and the penetration depth of gold for electrons, (7–8 nm at 1.5 eV). Experiments were conducted at room temperature in transmission geometry with an almost collinear arrangement between the pump and probe pulses. The background pressure in the experimental vacuum chamber was ~$10^{-9}$ mbar.

## SAMPLE PREPARATION

Gold nanoparticles coated with 1-octanethiol (C8), 1-decanethiol (C12), or 1-octanedecanethiol (C18) were synthesized [37] as follows: 0.125 mmol of chloro(triphenylphosphine)gold(I), 0.375 mmol of the corresponding alkanethiol, and 20 mL of benzene were mixed in a 100 mL round bottom flask at room temperature for 10 min and then introduced in a heating bath at 90°C until it reached reflux conditions. Then, 1.25 mmol of the reducing agent borane tert-butylamine complex dissolved in 20 mL of benzene was quickly added to the solution, left for reaction during 1h, and finally cooled at room temperature. The nanoparticle solution was precipitated with methanol and cleaned five times in acetone by centrifugation and redispersion in a sonication bath. The nanoparticles were dried in vacuum for storing and dissolved in toluene before use (for the case of C18-coated nanoparticles, they were slightly heated (~30 °C) to completely dissolve). 1-Octanethiol NPs: the analysis of the TEM image of Fig. 3a (main text) indicated a core diameter of 4.4 ± 0.5 nm and an interparticle distance of 6.0 ± 0.6 nm. 1-Dodecanethiol NPs: the analysis of the TEM image of Fig. 3b (main text) indicated a core diameter of 5.16 ± 0.58 nm with 9% polydispersity and an interparticle distance of 7.17 ± 0.68 nm. 1-Octadecanethiol NPs: the analysis of the TEM image of Fig. 3c (main text) indicated a core diameter of 4.8 ± 0.3 nm and an interparticle distance of 7.1 ± 0.5 nm. The TEM analysis was performed with the software package ImageJ. The projected area of the nanoparticles was calculated by standard particle analysis and default graylevel threshold. The diameter was calculated assuming spherical nanoparticles and averaging to all nanoparticles (>1000). The interparticle distance was obtained by the calculation of the center of mass of every nanoparticle and avereging to all nanoparticle neighbour pairs. The results from the TEM image analysis are consistent with the ones obtained with electron diffraction (Table 1, main text).

## CROSS-CORRELATION FUNCTIONS

Considering $I(s,\varphi)$ the diffracted intensity at the scattering vector $s$ and the azimuthal angle $\varphi$, the one-dimensional Fourier Transform (FT) over the $\varphi$ coordinate is defined as $F_\varphi\{I(s,\varphi)\}$. The Fourier transform of $I(s,\varphi)$ yields a spectrum where the most pronounced frequencies indicate the presence of a signal with a corresponding periodicity. The non-normalized cross-correlation function is obtained from the one-dimensional Fourier Transform as $C(\Delta) = \langle I(s,\varphi)I(s,\varphi+\Delta)\rangle_\varphi = Re\left(F_\varphi^{-1}\left(\left|F_\varphi\{I(s,\varphi)\}\right|^2\right)\right)$, where $\Delta$ is the angle shift (see Fig. 1 main text) and $\langle\ \rangle_\varphi$ denotes an averaging over $\varphi$. $C(\Delta)$ is then used to compute the normalized CCF:[23–26] $C_{norm}(\Delta) = \frac{\langle I(s,\varphi)I(s,\varphi+\Delta)\rangle_\varphi - \langle I(s,\varphi)\rangle^2_\varphi}{\langle I(s,\varphi)\rangle^2_\varphi}$. The local symmetry of the NPs, within each supracrystal, can be obtained by applying this methodology in a region of the electron diffraction patterns very close to the central beam (DC), at small-angle. The intensity profile $I(s_1,\varphi)$ of the scattering at $s_1$ and the Fourier Spectrum $F_\varphi\{I(s_1,\varphi)\}$ of the C8, C12, C18 samples are reported in Fig. S1. The CCFs reported in Fig. 5c (main text) are obtained from the diffraction patterns simulated as the squared amplitude of the Fourier Transform of the TEM images (Fig. 3, main text). In these patterns, the analysis is carried out at the first order of diffraction of $s_1$, indicated in the FTs in Fig. 3(b, d, f) in the main text. The corresponding $s_1$ values for the three samples are listed in Table1. In the experimental diffraction patterns, the diffraction at the first order of $s_1$ is overexposed by the DC. For this reason, the analysis at $s_1$ in the experiment (Fig. 1, main text) was carried out at the seventh diffraction order $s_1$ (~0.6-0.7 Å$^{-1}$). We point out that different diffraction orders from the same real-space object show an identical CCF periodicity, as demonstrated in [11].

## FOURIER TRANSFORMS OF TEM IMAGES AND SIMULATIONS

The TEM images and the simulations (Fig.3 and Fig. 4, main text) cover an area of 390 nm x 390 nm, estimated from scale-bars. The area in the TEM images is included within the one probed experimentally with UED. Artifacts in all two-dimensional Fourier transforms associated with image re-scaling were avoided by converting image to a binary (black and white) image, followed by multiplication with an apodizing cosine function to smooth the edges of the images and remove the scale-bar [38]. Each two-dimensional Fourier transform is normalized to the number of nanoparticles detected in the corresponding TEM or simulated image.

## AZIMUTHAL INTENSITY AND FOURIER SPECTRUM

The CCFs are computed from the azimuthal scattered intensity at $s_1$, $I(s_1, \varphi)$, through the the one-dimensional Fourier transform $F_\varphi\{I(s_1, \varphi)\}$ over the $\varphi$ coordinate (see Experimental Section, main text). In Fig. S1, the values of $I(s_1, \varphi)$ and $F_\varphi\{I(s_1, \varphi)\}$ obtained with electron diffraction (a-c) and for the Fourier transform of the TEM images (d-f) are compared for each sample: C8 (top), C12 (middle), C18 (bottom). From Fig. S1 it is clear that the results from UED and from the FT simulation are consistent: in C8 and C12 the presence of a recognizable ν=6 frequency above the noise level indicates an hexagonal order of NPs, as opposed to the C18 sample. We point out that the 6-fold signals obtained for C8 and C12 in the FT simulations are higher in intensity than the signal from UED for the following reason: the real space region observed with diffraction was at least comparable to the electron beam spotsize at the focus, ~160 µm. The diffraction patterns simulated as the squared amplitude of the Fourier transform of the TEM images, instead, were calculated from a real space area of each imaged area is 390 nm x 390 nm. Thus, the UED experiment probes a level of disorder in the supracrystal which is at least 3 orders of magnitude higher than the simulation. This directly reflects to the data obtained with UED being affected by signal-to-noise ratio differently with respect to the FT simulations. Finally, we remark that the 4-fold modulation observed in $F_\varphi\{I(s_1, \varphi)\}$ with UED is due to the substrate.

This has already been demonstrated in a detailed study reported in [11], where we were able to carefully disentangle modulations at $s_1$ originating solely from the scattering of electrons from the gold NPs and their ligands, from the ones stemming from the substrate. The background analysis was carried out comparing data from the C12 supracrystal, to data obtained from: (i) an amorphous carbon coated TEM copper grid and (ii) an empty copper TEM grid.

## C8 AND C12 SAMPLES TRANSIENT ANNEALING

A 14% and a 10% transient $I(s_1)$ suppression is reported in Fig. S2 (a, c) for C8 and C12, separately. In each panel, three time delays ($t_1 < t_0$, $t_2 \sim t_0$, $t_3 > t_0$) are selected to retrieve the cross-correlation functions. The CCFs at three time delays ($t_1$, $t_2$, $t_3$) at $s_1$ are displayed in Fig. S2(b) for C8 and Fig. S2(d) for C12. The amplitude decrease of the CCFs as a function of time is consistent with the dynamics in the system retrieved from the time traces, evidencing a photo-induced disorder. However, the six-fold CCFs are found to transiently become more evident as higher signal-to-noise ratio upon photo-excitation, despite decreasing in amplitude, evidencing annealing of the grains in both samples. Thus, local order in the grains arrangement is impulsively triggered for each supracystal, within the relevant time-scales.

## C18 SAMPLE TRANSIENT RESPONSE

The velocity of the $I(s_1)$ suppression is directly related to the coupling strength between the electronic and structural degrees of freedom in each supracrystal, which significantly differs with the ligand length. As demonstrated by the CCF static analysis (Fig. 2a, main text), the distribution of NPs in the C18 sample is the one of a polycrystalline phase with no recognizable local symmetry. Thus, we cannot identify a distinct diffraction feature to follow the NPs distribution temporal evolution. For this reason, the temporal dependence of the scattered intensity $I(s_1)$ for the C18 supracrystal is not reported in Fig. 2c (see main text).

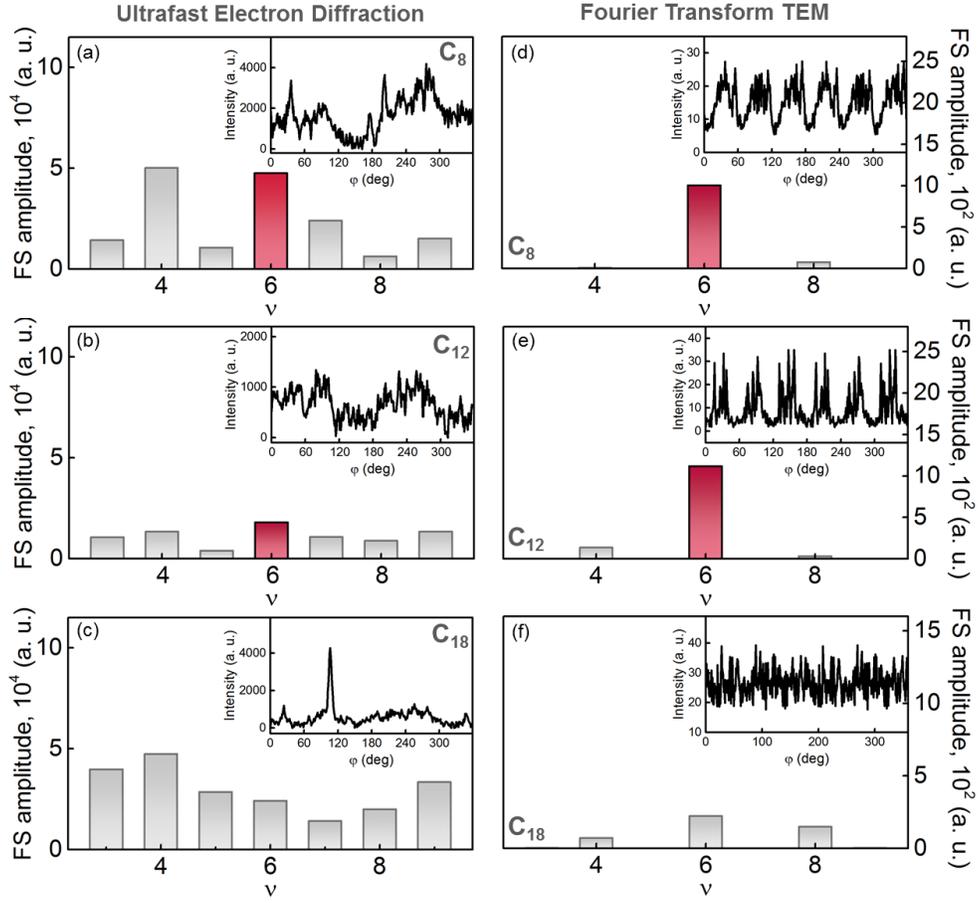

FIG. S1: Fourier Spectrum $F_\varphi\{I(s_1,\varphi)\}$ of the C8, C12, C18 samples. (a-c) $F_\varphi\{I(s_1,\varphi)\}$ from small-angle electron diffraction. (d-f) $F_\varphi\{I(s_1,\varphi)\}$ from the Fourier Transform of the TEM image. Insets in each panel: intensity profile $I(s_1,\varphi)$ of the scattering at $s_1$.

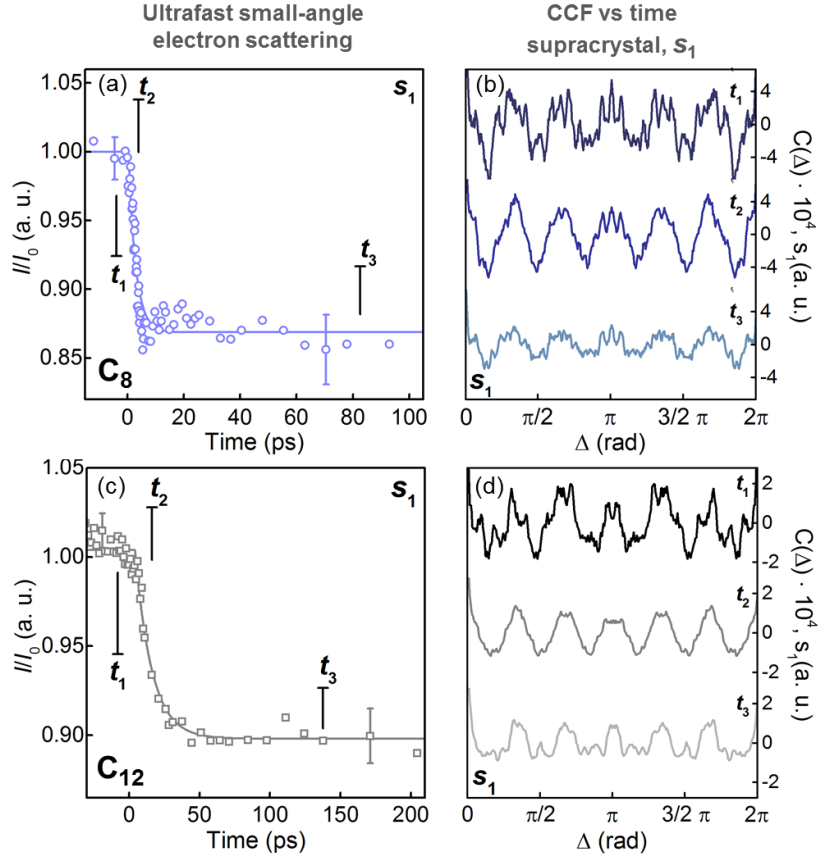

FIG. S2: Grains transient annealing in C8 and C12 samples. a,c) Dynamics of $I(s_1)$ for the (a) C8 and (c) C12 supracrystals. Each intensity data set is normalized to its average value before time zero ($I_0$) and it has been fitted to a mono-exponential function. b,d) CCFs at $s_1$ at three time delays ($t_1$, $t_2$, $t_3$) for (b) C8 and for (d) C12. The six-fold CCFs are found to transiently become more evident as higher signal-to-noise ratio upon photoexcitation, despite decreasing in amplitude, evidencing annealing of the grains in both samples. Data in (c, d) are adapted from our previous work [11].